\documentclass[twocolumn,showpacs,amsmath,amssymb,prd]{revtex4-1}

\usepackage{graphicx}
\usepackage{epsfig}
\usepackage{dcolumn}
\usepackage{bm}
\usepackage{mathrsfs,amsmath}

\def\beq{\begin{equation}}
\def\enq{\end{equation}}
\def\ba{\begin{eqnarray}}
\def\ea{\end{eqnarray}}
\def\<{\langle}
\def\>{\rangle}

\newcommand{\fr}{$f(R)$}
\newcommand{\Lam}{$\Lambda$CDM }
\newcommand{\lam}{$\lambda_1$ }

\urldef\mgcamb\url{http://www.sfu.ca/~gza5/MGCAMB.html}
\urldef\lensnoise\url{http://lesgourg.web.cern.ch/lesgourg/codes.html}

\begin{document}

\title{Constraining Modified Gravity with Euclid} \author{Matteo Martinelli$^1$, Erminia Calabrese$^1$, Francesco De Bernardis$^1$, Alessandro Melchiorri$^1$, Luca Pagano$^1$ and Roberto Scaramella$^2$}
\affiliation{$^1$Physics Department and INFN, Universita' di Roma ``La Sapienza'', Ple Aldo Moro 2, 00185, Rome, Italy}
\affiliation{$^2$INAF, Osservatorio Astronomico di Roma, via Frascati 33, 0040  Monte Porzio Catone (RM), Italy}

\begin{abstract}
Future proposed satellite missions as Euclid can offer the opportunity to test general relativity on cosmic scales
through mapping of the galaxy weak lensing signal. In this paper we forecast the ability of these experiments to
constrain modified gravity scenarios as those predicted by scalar-tensor and $f(R)$ theories. We found that Euclid will
improve constraints expected from the PLANCK satellite on these modified gravity models by two orders of magnitude.
We discuss parameter degeneracies and the possible biases introduced by modified gravity.
\end{abstract}


\date{\today}
\maketitle
\section{Introduction}


Understanding the nature of the current observed accelerated expansion of our universe
is probably the major goal of modern cosmology. Two possible mechanisms can
be at work: either our Universe is described by general relativity (GR, hereafter) and its energy 
content is dominated by a negative pressure component, coined "dark energy", either
only "standard" forms of matter exist and GR is not valid on cosmic scales 
(see e.g. \cite{Tsujikawa:2010sc}, \cite{Silvestri:2009hh}).

All current cosmological data are consistent with the choice of a cosmological
constant as dark energy component with equation of state $w=P/\rho=-1$ where $P$ and $\rho$
are the dark energy pressure and density respectively 
(see e.g. \cite{Serra:2009yp}, \cite{Padmanabhan:2002ji}, \cite{Peebles:2002gy}). 

While deviations at the level of $\sim 10 \%$ on $w$ assumed as constant 
are still compatible with observations and bounds on $w$ are even weaker if 
$w$ is assumed to be redshift-dependent, it may well be that future measurements
will be unable to significantly rule out the cosmological constant value of $w=-1$.

Measuring $w$, however, is just part of the story. While the background
expansion of the universe will be identical to the one expected in the case of
a cosmological constant, the growth of structures with time could
be significantly different if GR is violated. Modified gravity models have recently been
proposed where the expansion of the universe is identical to the one produced
by a cosmological constant, but where the primordial perturbations that will
result in the large scale structures in the universe we observed today, grow
at a different rate (see e.g. \cite{Laszlo:2007td}, \cite{Amendola:2007rr}, \cite{Linder:2007hg}).

Weak lensing measurements offer the great opportunity to map the growth
of perturbations since they relate directly to the dark matter distribution
and are not plagued by galaxy luminous bias (\cite{Bartelmann:1999yn}, \cite{Refregier:2003ct}, \cite{VanWaerbeke:2003uq}).
Recent works have indeed make use of current weak lensing measurements, combined
with other cosmological observables, to constrain modified gravity
yielding no indications for deviations from GR 
(\cite{Daniel:2010ky}, \cite{Lombriser:2010mp}, \cite{Bean:2010zq}, \cite{Zhao:2010dz}, 
\cite{Daniel:2008et}, \cite{Daniel:2009kr}). 

Next proposed satellite mission as Euclid (\cite{Refregier:2010ss}, \cite{Refregier:2006vt}) 
or WFIRST \cite{Gehrels:2010fn} could measure the galaxy weak
lensing signal to high precision, providing a detailed history of structure
formation and the possibility to test GR on cosmic scales.

In this paper we study the ability of these future satellite missions to
constrain modified gravity models and to possibly falsify a cosmological 
constant scenario. Respect to recent papers that have analyzed this possibility 
(e.g. \cite{Thomas:2008tp}, \cite{Xia:2009gb})
 we improve on several aspects. First of all, we forecast the future constraints
by making use of Monte Carlo simulations on synthetic realisations of datasets.
Previous analyses (see e.g. \cite{Amendola:2007rr}, \cite{Zhao:2008bn}, \cite{Heavens:2007ka}) 
often used the Fisher matrix formalism that, while
fast, it may lose its reliability when Gaussianity is not respected 
due, for instance, to strong parameter degeneracies.
Secondly we properly include the future constraints achievable by the
Planck satellite experiment, also considering CMB lensing, that is
a sensitive probe of modified gravity (see e.g. \cite{Serra:2009kp},
\cite{Calabrese:2009tt} and references therein).
Thirdly, we discuss the parameter degeneracies and the impact of
modified gravity on the determination of cosmological parameters,.
Finally we focus on $f(R)$ and scalar-tensor theories, using the 
general parametrization proposed by \cite{Zhao:2008bn}.

Our paper is structured as follows. In Section \ref{sec:ii} 
we introduce the parametrization used to describe departures from GR, and then specialize to the case of \fr \ 
and scalar-tensor theories. In Section \ref{sec:iii} we describe Galaxy weak-lensing, while in
section \ref{sec:iv} we discuss how to extract lensing information from CMB data. 
We review the analysis method and the data forecasting in Section \ref{sec:v}. 
In Section \ref{sec:vi} we present our results and we derive our conclusions in Section \ref{sec:vii}.

\section{Modified gravity parametrization}\label{sec:ii}
In this section we describe the formalism we use to parametrize departures from general relativity.

\subsection {Background expansion}

In our analysis we fix the background expansion to a standard $\Lambda$CDM cosmological model.
The reasons for this choice are multiple; $\Lambda$CDM is currently the best
fit to available data and popular models of modified gravity, e.g. $f(R)$, closely mimic $\Lambda$CDM at the background level
with differences which are typically smaller than
the precision achievable with geometric tests~\cite{Hu:2007nk}. The most significant departures happen at the level of
growth of structure and, by restricting ourselves to $\Lambda$CDM backgrounds, we are able isolate them.

\subsection {Structure formation}

In modified gravity models we expect departures from the standard growth of structure,
even when the expansion history matches exactly the $\Lambda$CDM one. Dark matter clustering, as well as the evolution
of the metric potentials, is changed and can be scale-dependent. Moreover, typically there might be an effective anisotropic stress introduced by the modifications and the two potentials appearing in the metric element, $\Phi$ and $\Psi$, are not necessarily equal, as is in the $\Lambda$CDM model.
Here we focus  on the effect of the modified evolution of the potential, $\Phi+\Psi$, on the CMB power spectra.\\ 
In order to study the potentials
evolution and to evaluate the growth of perturbations in modified gravity models we employ the MGCAMB code developed in~\cite{Zhao:2008bn} (and publicly available at \mgcamb) . In this code the modifications to the Poisson and anisotropy equations are parametrized by two functions $\mu(a,k)$ and $\gamma(a,k)$ defined by: 
\beq\label{mu}
k^2\Psi=-\frac{a^2}{2M_P^2}\mu(a,k)\rho\Delta \ ,\\
\enq
\beq\label{gamma}
\frac{\Phi}{\Psi}=\gamma(a,k) \,,
\enq
 where $\rho\Delta\equiv\rho\delta+3\frac {aH}{k}(\rho+P)v$ is the comoving density perturbation.
These functions can be expressed using the parametrization introduced by~\cite{Bertschinger:2008zb} (and used in~\cite{Zhao:2008bn}):
\beq\label{par_mu_old}
\mu(a,k)=\frac{1+\beta_1\lambda_1^2\,k^2a^s}{1+\lambda_1^2\,k^2a^s}\,,\\
\enq
\beq\label{par_gamma}
\gamma(a,k)=\frac{1+\beta_2\lambda_2^2\,k^2a^s}{1+\lambda_2^2\,k^2a^s}\,,
\enq
where the parameters $\beta_i$ can be thought of as dimensionless couplings, $\lambda_i$ as dimensionful
length scales and $s$ is determined by the time evolution of the characteristic length scale of the theory.
$\Lambda$CDM cosmology is recovered for $\beta_{1,2}=1$ or $\lambda_{1,2}^2=0$ \rm {Mpc}$^{2}$.

\subsubsection{Scalar-Tensor theories}

This parametrization can be to constrain chameleon type scalar-tensor theories, where the gravity Lagrangian
is modified with the introduction of a scalar field~\cite{Khoury:2003aq}.  As shown in~\cite{Zhao:2008bn},
for this kind  of theories the parameters $\{\beta_i,\lambda_i^2\}$ are related in the following way:
\beq\label{ST_relation}
\beta_1=\frac{\lambda_2^2}{\lambda_1^2}=2-\beta_2\frac{\lambda_2^2}{\lambda_1^2}
\enq
and $1\lesssim s\lesssim4$.\\ 
This implies that we can analyze scalar-tensor theories adding 3 independent parameters to the standard 
cosmological parameter set.

\subsubsection{\fr \  theories}
In the specific case of $f(R)$ theories we can additionally reduce the number
of free parameters since $f(R)$ theories correspond to a fixed coupling $\beta_1=4/3$~\cite{Magnano:1993bd}.
Moreover, to have $\Lambda$CDM background expansion the $s$ parameter must be $\sim 4$~\cite{Zhao:2008bn}. 
The parametrization in Eq.~(\ref{par_mu_old}) effectively neglects a factor representing the rescaling of the Newton's constant 
(e.g. $(1+f_R)^{-1}$ in $f(R)$ theories) that, as pointed out in \cite{Giannantonio:2009gi}, is very close to unity in models that satisfy 
local tests of gravity~\cite{Hu:2007nk} and so negligible. 
However, when studying the $f(R)$ case, we need to include it to get a more precise MCMC analysis (see \cite{Giannantonio:2009gi}
for the detailed expression of Eq.~(\ref{par_mu_old})).
Even with this extended parametrization,  we have only one free parameter left, the length scale $\lambda_1$. In this work we will constrain $f(R)$
theories through this parameter, evaluating the effects of these theories on gravitational lensing.

\section{Galaxy weak Lensing} \label{sec:iii}
Being sensitive the the growth rate of the structure, weak lensing can be very useful to constrain modified gravity and to distinguish between
 various modified gravity models when combined with CMB observations.\\More generally, weak lensing (see \cite{Munshi:2006fn} for a recent review or {\tt
http://www.gravitationallensing.net}) is a particularly powerful
probe for Cosmology, since it simultaneously measures the growth of
structure through the matter power spectrum, and the geometry of the
Universe through the lensing effect. Since weak lensing probes the
dark matter power spectrum directly, it is not limited by any
assumption about the galaxy bias (how galaxies are
clustered with respect to the dark matter) that represents one of the main limitations of galaxy surveys.\\
Following \cite{Bartelmann:1999yn} one can describe the distortion of the images of distant galaxies through the tensor:\\
\begin{center}
\(\psi_{ij}=\left(
\begin{array}{cc}
-\kappa-\gamma_1 & -\gamma_2 \\
-\gamma_2 & -\kappa+\gamma_2 \\
\end{array}\right)\)
\end{center}
where $\kappa$ and $(\gamma_1,\gamma_2)$ represents respectively the
convergence (or magnification) and the shear (or stretching)
component of the distortion. The reconstruction of matter density
field can be conducted by looking at the correlations of the image
distortions. The observable one has to deal with, will be hence a
convergence power spectra \cite{kai:1992,Kaiser:1996tp}:
\begin{equation}\label{convergencegalaxy}
 P_{jk}(\ell)=H_0^3\int_0^\infty\frac{dz}{E(z)}W_i(z)W_j(z)P_{NL}[P_L\left(\frac{H_0\ell}{r(z)},z\right)]
\end{equation}
where $P_{NL}$ is the non-linear matter power spectrum at redshift
$z$, obtained correcting the linear one $P_{L}$. $W(z)$ is a weighting function, with subscripts $i$ and $j$
indicating the bins in redshifts. The function $W(z)$ also encodes the
cosmological information, being:
\begin{equation}\label{weight}
    W_i(z)=\frac{3}{2}\Omega_m(1+z)\int_{z_i}^{z_{i+1}}dz^\prime\frac{n_i(z^\prime)r(z,z^\prime)}{r(0,z^\prime)}
\end{equation}
where:$$r(z,z^\prime)=\int_z^{z^\prime}\frac{dz^\prime}{E(z^\prime)}$$with
$E(z)=H(z)/H_0$ and $n_i(z^\prime)$ is the fraction of sources
belonging to the $i-th$ bin.\\ The observed convergence power
spectra is affected mainly by a systematic arising from the
intrinsic shear of galaxies $\gamma^2_{rms}$. This uncertainties can
be reduced averaging over a large number of sources. The observed
convergence power spectra will be hence:
\begin{equation}\label{obsconv}
    C_{jk}=P_{jk}+\delta_{jk}\gamma^2_{rms}\tilde{n}_j^{-1}
\end{equation}
where $\tilde{n}_j$ is the number of sources per steradian in the
$j-th$ bin.


\section{CMB lensing extraction} \label{sec:iv}

In the analysis we perform, we choose to introduce, in addition to galaxy weak lensing, the information derived from CMB lensing extraction.\\
Gravitational CMB lensing, as already shown in Ref.~\cite{Perotto:2006rj}, can improve significantly the CMB constraints on several cosmological parameters, 
since it is strongly connected with the growth of perturbations and gravitational
potentials at redshifts $z < 1$ and, therefore, it can break important degeneracies.
The lensing deflection field $d$ can be related to the lensing potential $\phi$ as $d=\nabla\phi$~\cite{hirata:2003}. In harmonic space, the deflection 
and lensing potential multipoles follows:
\begin{equation}
d_\ell^m=-i\sqrt{\ell(\ell+1)}\phi_\ell^m, \label{eq:defd}
\end{equation}
\noindent and therefore, the power spectra
$C^{dd}_\ell\equiv\left\langle d_\ell^m d_\ell^{m*}\right\rangle$
and
$C_\ell^{\phi\phi}\equiv\left\langle\phi_\ell^m\phi_\ell^{m*}\right\rangle$
are related through:
\begin{equation}
C_\ell^{dd}=\ell(\ell+1)C_\ell^{\phi\phi}.
\end{equation}

Gravitational lensing introduces a correlation between different CMB multipoles (that otherwise would be fully uncorrelated) through the relation:
\begin{equation}
\left\langle a_\ell^m
b_{\ell'}^{m'}\right\rangle=(-1)^m\delta_m^{m'}\delta_\ell^{\ell'}C_\ell^{ab}+
\sum_{LM}{\Xi^{mm'M}_{\ell\ \ell'\ L}\phi^M_L}~,
\label{eq:lens_corr}
\end{equation}
\noindent where $a$ and $b$ are the ${T,E,B}$ modes and $\Xi$ is a linear combination of the unlensed
power spectra $\tilde{C}_\ell^{ab}$ (see \cite{lensextr} for details).\\
In order to obtain the deflection power spectrum from the observed
$C_\ell^{ab}$, we have to invert Eq.~(\ref{eq:lens_corr}), defining
a quadratic estimator for the deflection field given by:
\begin{equation}
d(a,b)_L^M=n_L^{ab}\sum_{\ell\ell'mm'}W(a,b)_{\ell\ \ell'\
L}^{mm'M}a^m_\ell b^{m'}_{\ell'}~, \label{eq:estimator}
\end{equation}
\noindent where $n_L^{ab}$ is a normalization factor needed to construct an unbiased estimator ($d(a,b)$ must satisfy Eq.~(\ref{eq:defd})).
This estimator has a variance:
\begin{equation}
 \langle d(a,b)_L^{M*} d(a',b')_{L'}^{M'}\rangle\equiv \delta_{L}^{L'}\delta^{M'}_{M}(C_L^{dd}+N_L^{aa'bb'})
\end{equation}
that depends on the choice of the weighting factor $W$
and leads to a noise $N_L^{aa'bb'}$ on the deflection power spectrum $C_L^{dd}$ obtained through this method. The choice of $W$ and the particular lensing estimator we employ will be described in the next section. \\


\section{Future data analysis} \label{sec:v}

\subsection{Galaxy weak lensing data}
\begin{table}[h]
\begin{center}
\begin{tabular}{cccccccc}
$n_{gal} (arcmin^{-2})$ \hspace{5pt} & $redshift$\hspace{5pt}  &$f_{sky}$\hspace{5pt}  & $\gamma^2_{rms}$\\
\hline &&&&&&\\
$35$\hspace{5pt}&$0<z<2$\hspace{5pt} &$0.5$\hspace{5pt} &$0.22$\hspace{5pt}\\
\hline \hline
\end{tabular}
\caption{Specifications for the Euclid like survey considered in
this paper. The table shows the number of galaxies per square
arcminute ($n_{gal}$), redshift range, $f_{sky}$ and intrinsic shear
($\gamma^2_{rms}$).} \label{tabeuclid}
\end{center}
\end{table}

Future weak lensing surveys will measure photometric redshifts of
billions of galaxies allowing the possibility of 3D weak lensing
analysis
(e.g.\cite{Heavens:2003,Castro:2005,Heavens:2006,Kitching:2007}) or
a tomographic reconstruction of growth of structures as a function
of time through a binning of the redshift distribution of galaxies,
with a considerable gain of cosmological information (e.g. on
neutrinos \cite{Hannestad:2006as}; dark energy \cite{Kitching:2007};
the growth of structure \cite{art:Bacon,art:Massey}  and map the
dark matter distribution as a function of redshift
\cite{art:Taylor}).\\Here we use typical specifications for futures
weak lensing surveys like the Euclid experiment, observing about $35$ galaxies per square arcminute in the redshift range $0<z<2$ with an uncertainty of about $\sigma_z=0.03(1+z)$
(see \cite{Refregier:2006vt}), to build a mock dataset of convergence power
spectra. Table \ref{tabeuclid} shows the number of galaxies per
arcminute$ ^{-2}$ ($n_{gal}$), redshift range, $f_{sky}$ and
intrinsic shear for this survey. The expected $1\sigma$ uncertainty
on the convergence power spectra $P(\ell)$ is given by
\cite{Cooray:1999rv}:
\begin{equation}\label{sigmaconv}
    \sigma_{\ell}=\sqrt{\frac{2}{(2\ell+1)f_{sky}\Delta_{\ell}}}\left(P(\ell)+\frac{\gamma_{rms}^2}{n_{gal}}\right)
\end{equation}
For the convergence power spectra we use $\ell_{max}=1500$ in order
to exclude the scales where the non-linear growth of structure is more relevant
and the shape of the non-linear matter power spectra is, as a
consequence, more uncertain (see \cite{Smith:2002dz}). We calculate
the power spectra at a mean redshift $z=1$. $\Delta_{\ell}$ in the
(\ref{sigmaconv}) is the bin used to generate data. Here we choose
$\Delta_\ell=1$ for the range $2<\ell<100$ and $\Delta_\ell=40$ for
$100<\ell<1500$.\\In this first-order analysis we are not considering other systematic effects
as intrinsic alignments of galaxies, selection effects and shear measurements
errors due to uncertainties in the point spread function (PSF) determination. Of course future real data analysis will
require the complete treatment of these effects in order to avoid biases on the cosmological parameters.

\subsection{CMB data}

We create a full mock CMB datasets (temperature, E--polarization
mode and lensing deflection field) with noise properties consistent
with the Planck~\cite{:2006uk} experiment (see Tab.~\ref{tab:exp}
for specifications).

\begin{table}[!htb]
\begin{center}
\begin{tabular}{rccc}
Experiment & Channel & FWHM & $\Delta T/T$ \\
\hline
Planck & 70 & 14' & 4.7\\
\phantom{Planck} & 100 & 10' & 2.5\\
\phantom{Planck} & 143 & 7.1'& 2.2\\

$f_{sky}=0.85$ & & & \\
\hline
\hline
\end{tabular}
\caption{Planck experimental specifications. Channel frequency is given in GHz, FWHM (Full-Width at Half-Maximum) in arc-minutes, and the temperature 
sensitivity per pixel in $\mu K/K$. The polarization sensitivity is $\Delta E/E=\Delta B/B= \sqrt{2}\Delta T/T$.}

\label{tab:exp}
\end{center}
\end{table}

We consider for each channel a detector noise of $w^{-1} =
(\theta\sigma)^2$, where $\theta$ is the FWHM (Full-Width at
Half-Maximum) of the beam assuming a Gaussian profile and $\sigma$ is
the temperature sensitivity $\Delta T$ (see Tab.~\ref{tab:exp} for the polarization sensitivity). We therefore add to each $C_\ell$ fiducial spectra a 
noise spectrum given by:
\begin{equation}
N_\ell = w^{-1}\exp(\ell(\ell+1)/\ell_b^2) \, ,
\end{equation}
where $\ell_b$ is given by $\ell_b \equiv \sqrt{8\ln2}/\theta$.\\

In this work, we use the method presented in \cite{lensextr} to
construct the weighting factor $W$ of Eq.~(\ref{eq:estimator}). In
that paper, the authors choose $W$ to be a function of the power
spectra $C_\ell^{ab}$, which include both CMB lensing and primary
anisotropy contributions. This choice leads to five quadratic
estimators, with $ab={TT,TE,EE,EB,TB}$; the $BB$ case is excluded
because the method of Ref.~\cite{lensextr} is only valid when the
lensing contribution is negligible compared to the primary
anisotropy, assumption that fails for the B modes in the case of Planck. \\
The five quadratic estimators can be combined into a minimum
variance estimator which provides the noise on the deflection 
field power spectrum $C_\ell^{dd}$:
\begin{equation}
N_\ell^{dd}=\frac{1}{\sum_{aa'bb'}{(N_\ell^{aba'b'})^{-1}}}~.
\end{equation}
We compute the minimum variance lensing noise for Planck experiment by means of a routine publicly available at \lensnoise.
The datasets (which include the lensing deflection power spectrum) are analyzed with a full-sky exact likelihood routine available at the same URL.

\subsection{Analysis method}

In this paper we perform two different analysis. First, we evaluate
the achievable constraints on the $f(R)$ parameter $\lambda_1^2$ and
on the more general scalar-tensor parametrization including 
also $\beta_1$ and $s$. Secondly, we investigate the
effects of a wrong assumption about the modified gravity on the
cosmological parameters, by generating an $f(R)$ datasets with non-zero
$\lambda_1^2$ fiducial value and analysing it fixing $\lambda_1^2=0$ \rm {Mpc}$^{2}$. 
We conduct a full Monte Carlo Markov Chain analysis based on the publicly available
package \texttt{cosmomc} \cite{Lewis:2002ah} with a convergence
diagnostic using the Gelman and Rubin statistics.\\We sample the
following  set of cosmological parameters, adopting flat priors on
them: the baryon and cold dark matter densities $\Omega_{b}h^2$ and
$\Omega_{c}h^2$, the ratio of the sound horizon to the angular
diameter distance at decoupling $\theta_s$, the scalar spectral
index $n_s$, the overall normalization of the spectrum $A_s$ at
$k=0.002$ {\rm Mpc}$^{-1}$, the optical depth to reionization
$\tau$, and, finally, the modified gravity parameters $\lambda_1^2$,
$\beta_1$ and $s$.\\The fiducial model for the standard cosmological parameters
is the best-fit from the WMAP seven years analysis of
Ref.~\cite{wmap7} with $\Omega_{b}h^2=0.02258$, $\Omega_{c}h^2=
0.1109$, $n_s=0.963$, $\tau=0.088$, $A_s=2.43\times10^{-9}$,
$\Theta=1.0388$.\\ For modified gravity parameters, 
we first assume a fiducial value $\lambda_1^2=0$ \rm {Mpc}$^{2}$ and fix $\beta_1=1.33$ 
and $s=4$ to test the constraints achievable on the $f(R)$ model. We then 
repeat the analysis allowing $\beta_1$ and $s$ to vary. 
Furthermore, to investigate the ability of the combination of
Planck and Euclid data to detect an hypothetical modified gravity scenario, 
we study a model with fiducial $\lambda_1^2=300$ \rm {Mpc}$^{2}$ 
leaving $\lambda_1^2$, $\beta_1$ and $s$ as free variable parameters allowing them 
to vary in the ranges $0\leq\lambda_1^2\leq10^6$, $0.1\leq\beta_1\leq2$ and $1\leq s\leq4$.
Finally, we analyse a dataset with a fiducial value
$\lambda_1^2=300$ \rm {Mpc}$^{2}$ but assuming a \textit{wrong} $\Lambda$CDM model
 fixing $\lambda_1^2=0$ \rm {Mpc}$^{2}$, to investigate the bias introduced on
the cosmological parameter due to a wrong assumption about the
gravity model.

\section{Results} \label{sec:vi}

\begin{figure*}[h!]
\centering
\begin{tabular}{cc}
\epsfig{file=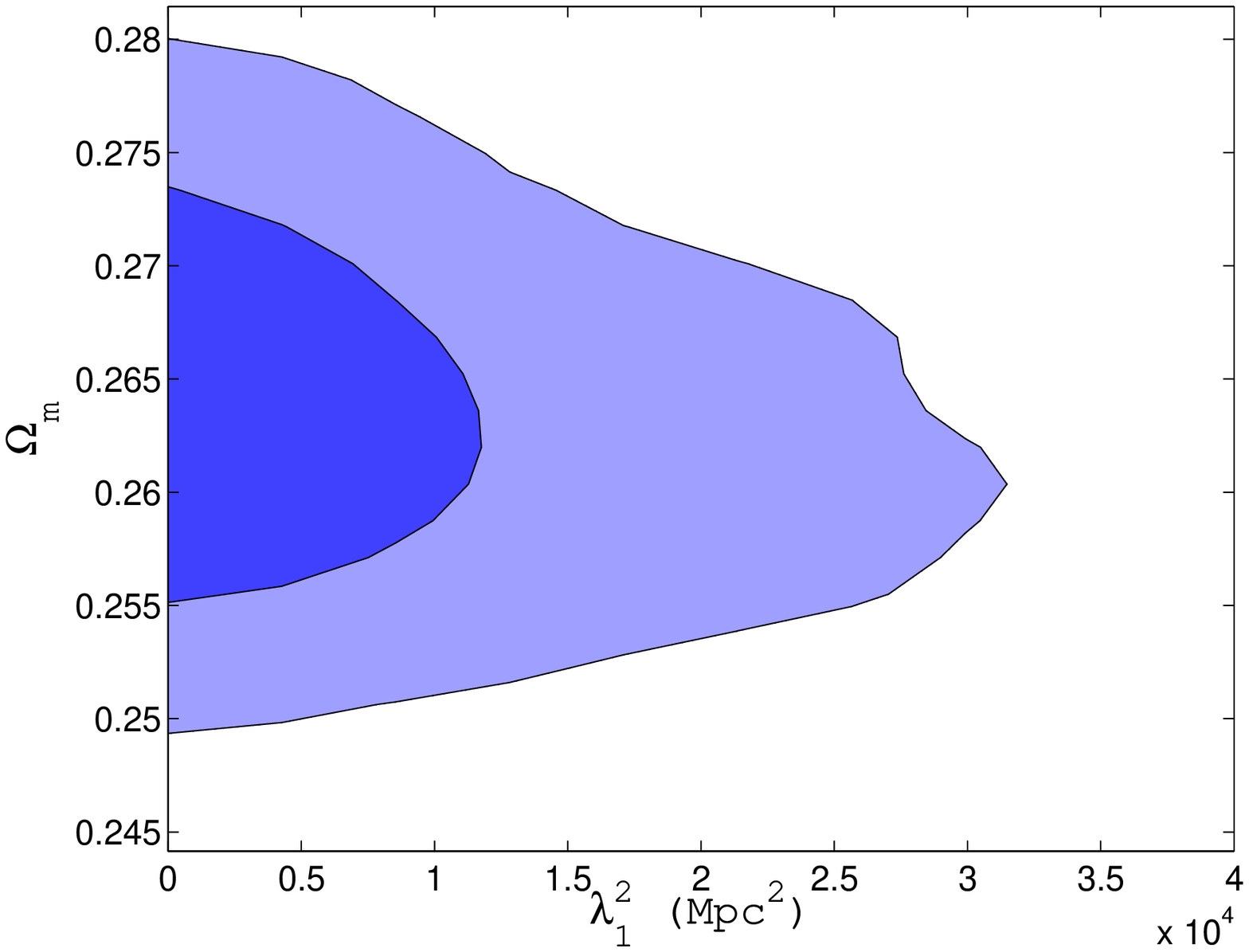,width=0.4\linewidth,clip=} &
\epsfig{file=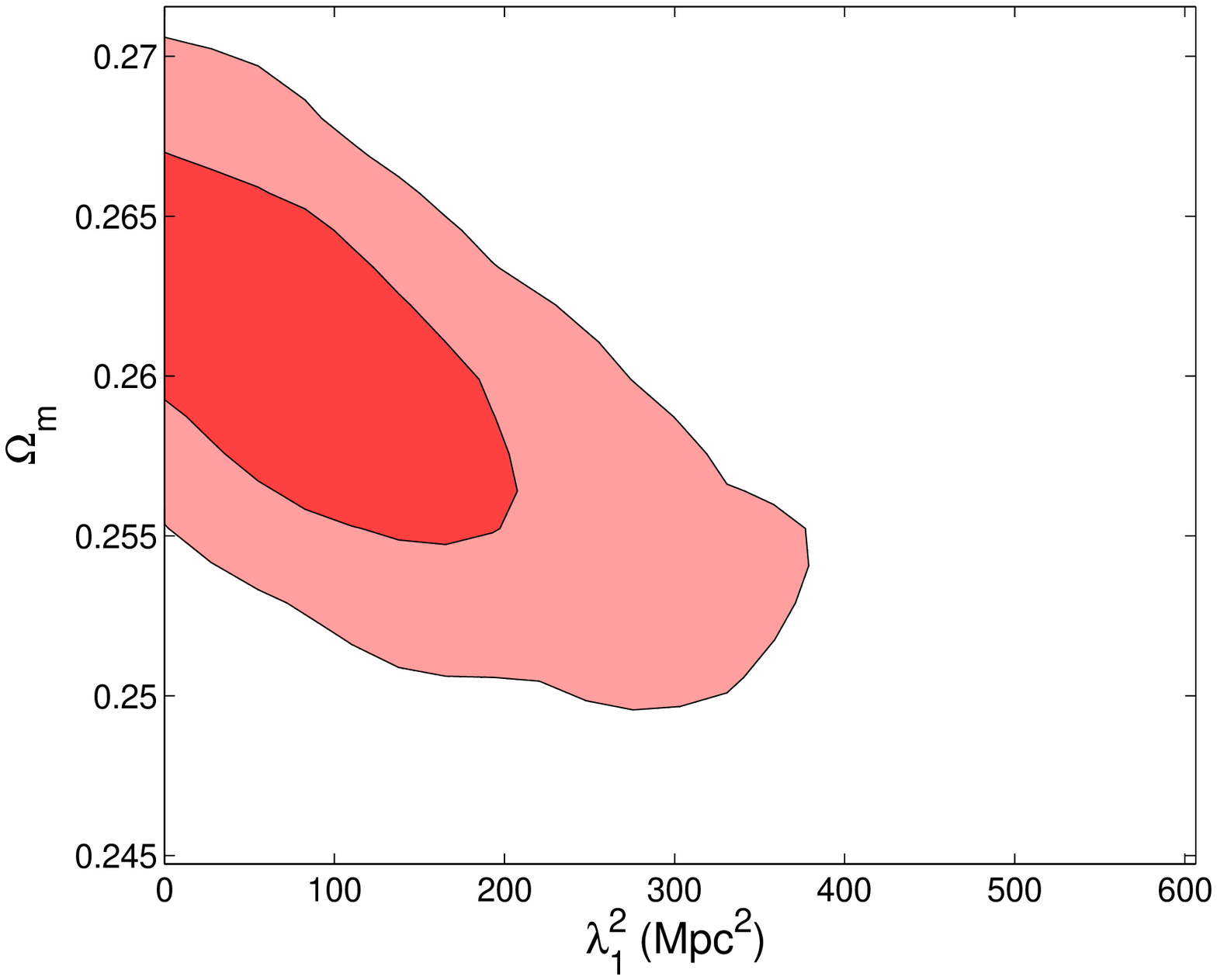,width=0.4\linewidth,clip=} \\
\epsfig{file=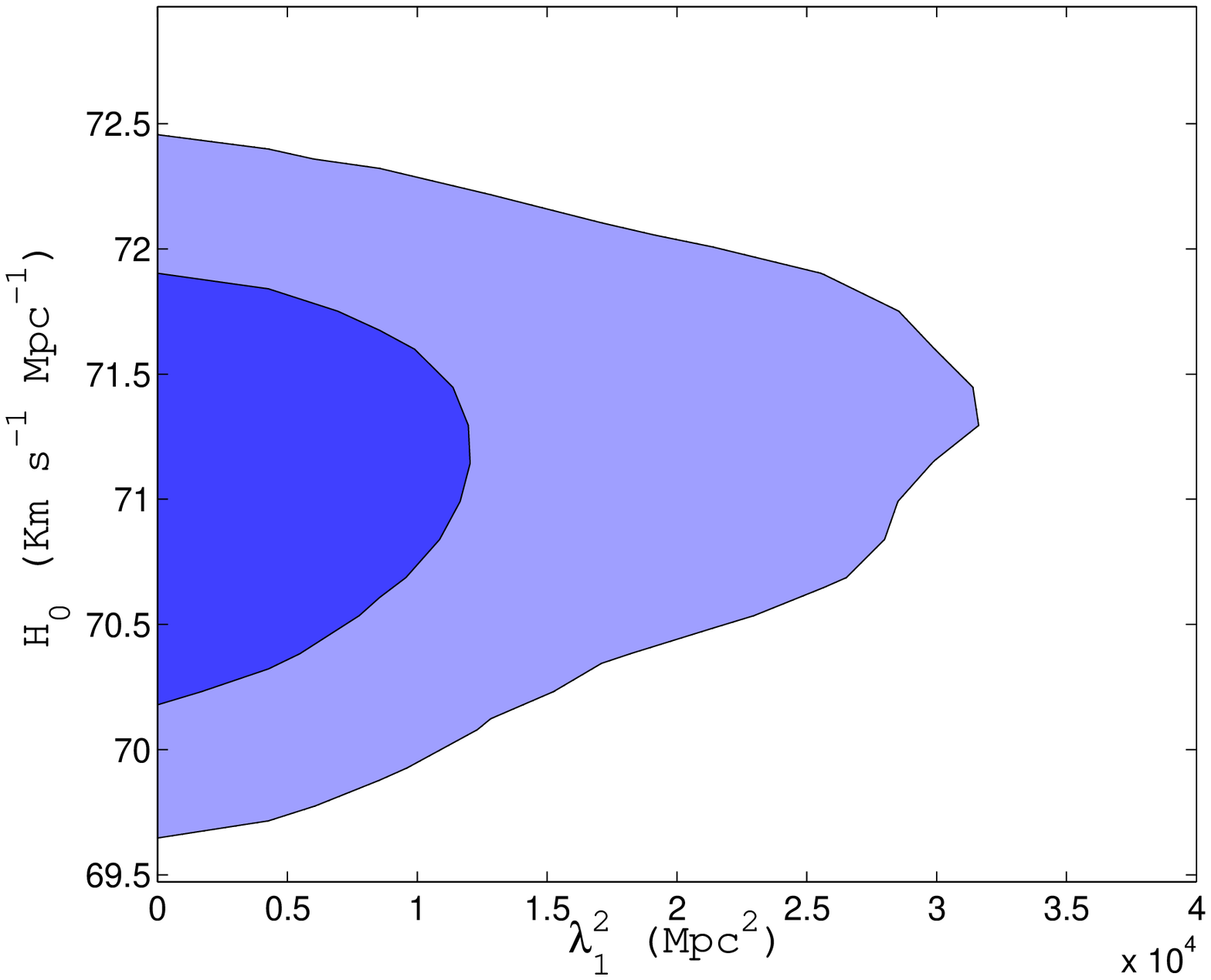,width=0.4\linewidth,clip=} &
\epsfig{file=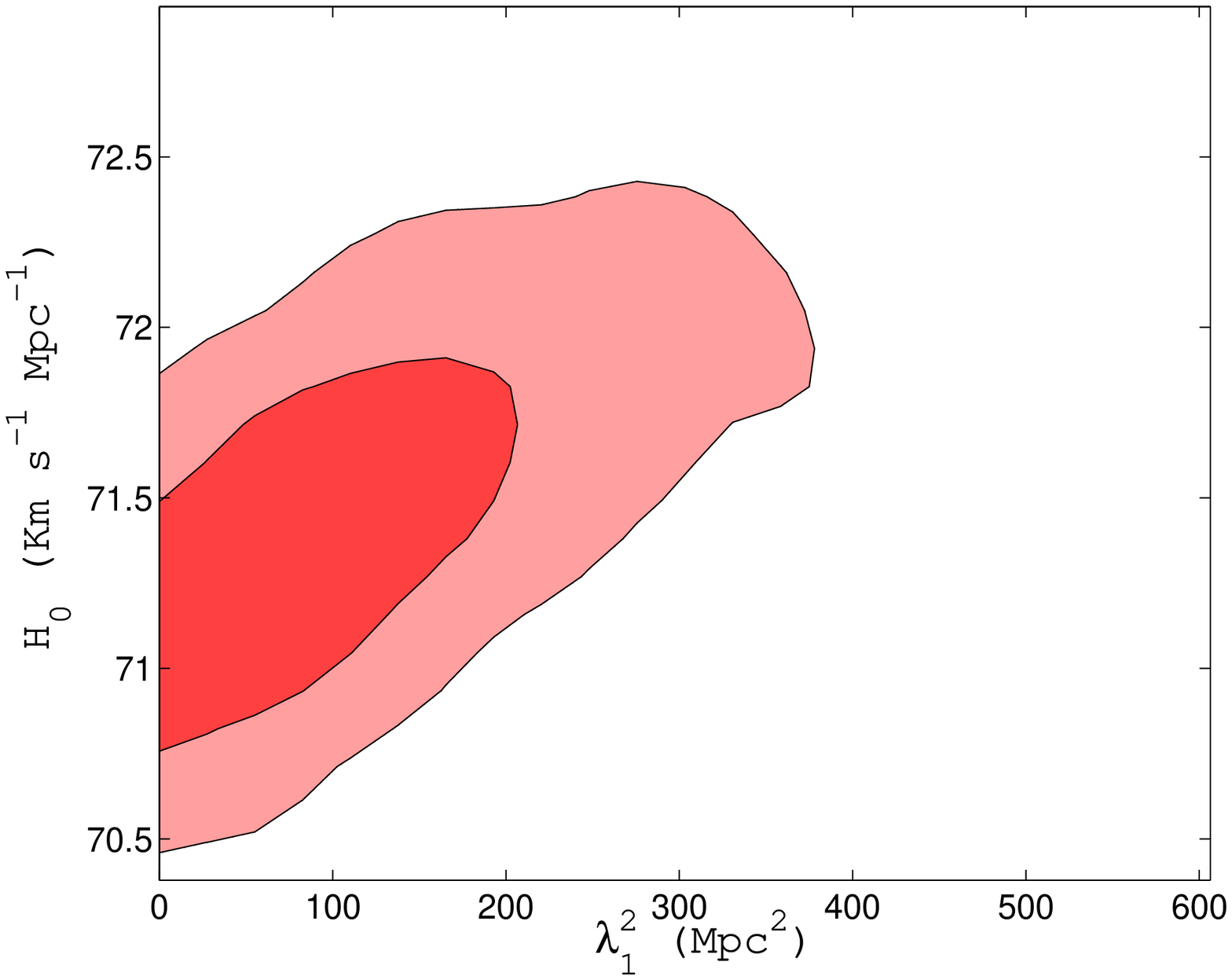,width=0.4\linewidth,clip=} \\
\epsfig{file=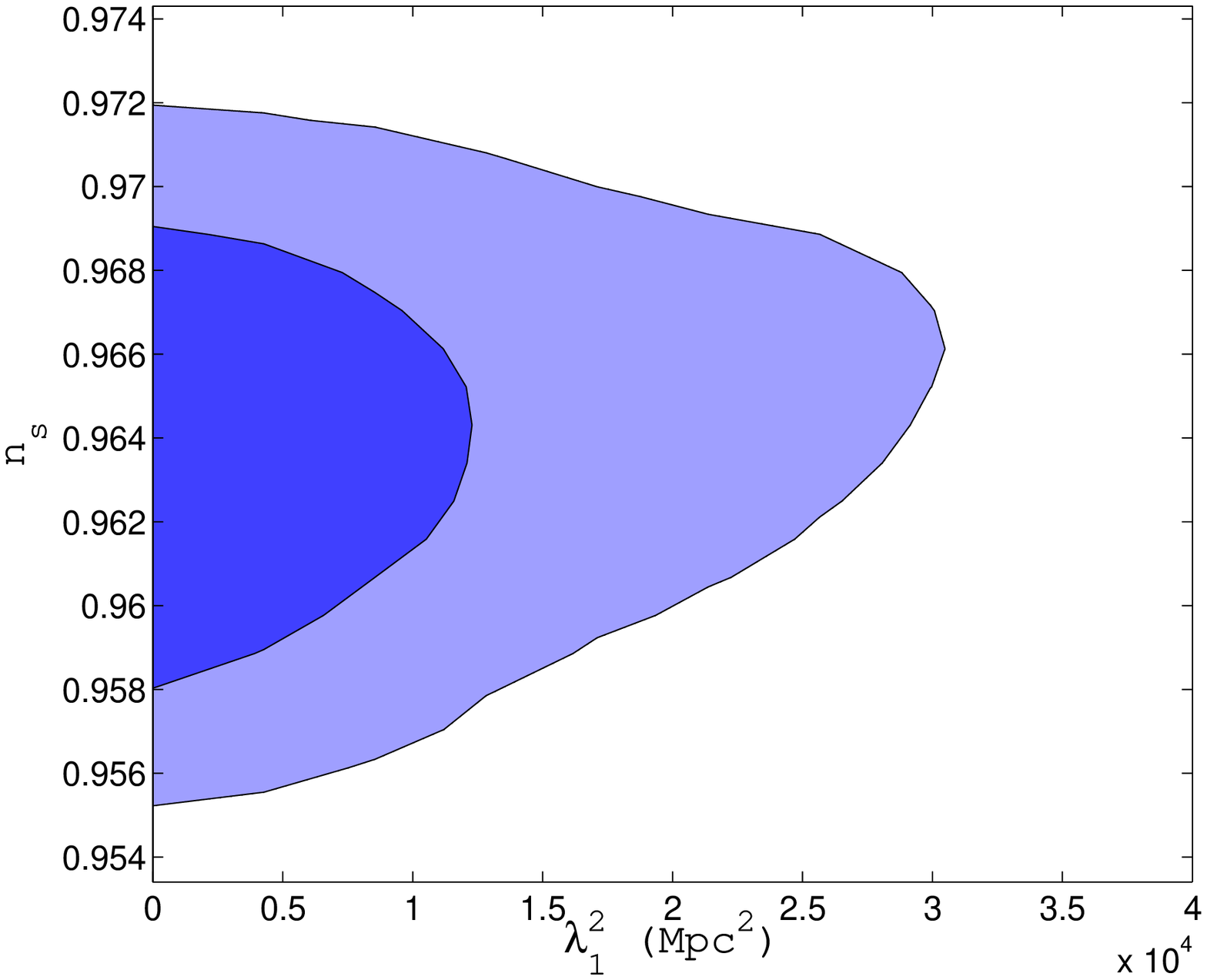,width=0.4\linewidth,clip=} &
\epsfig{file=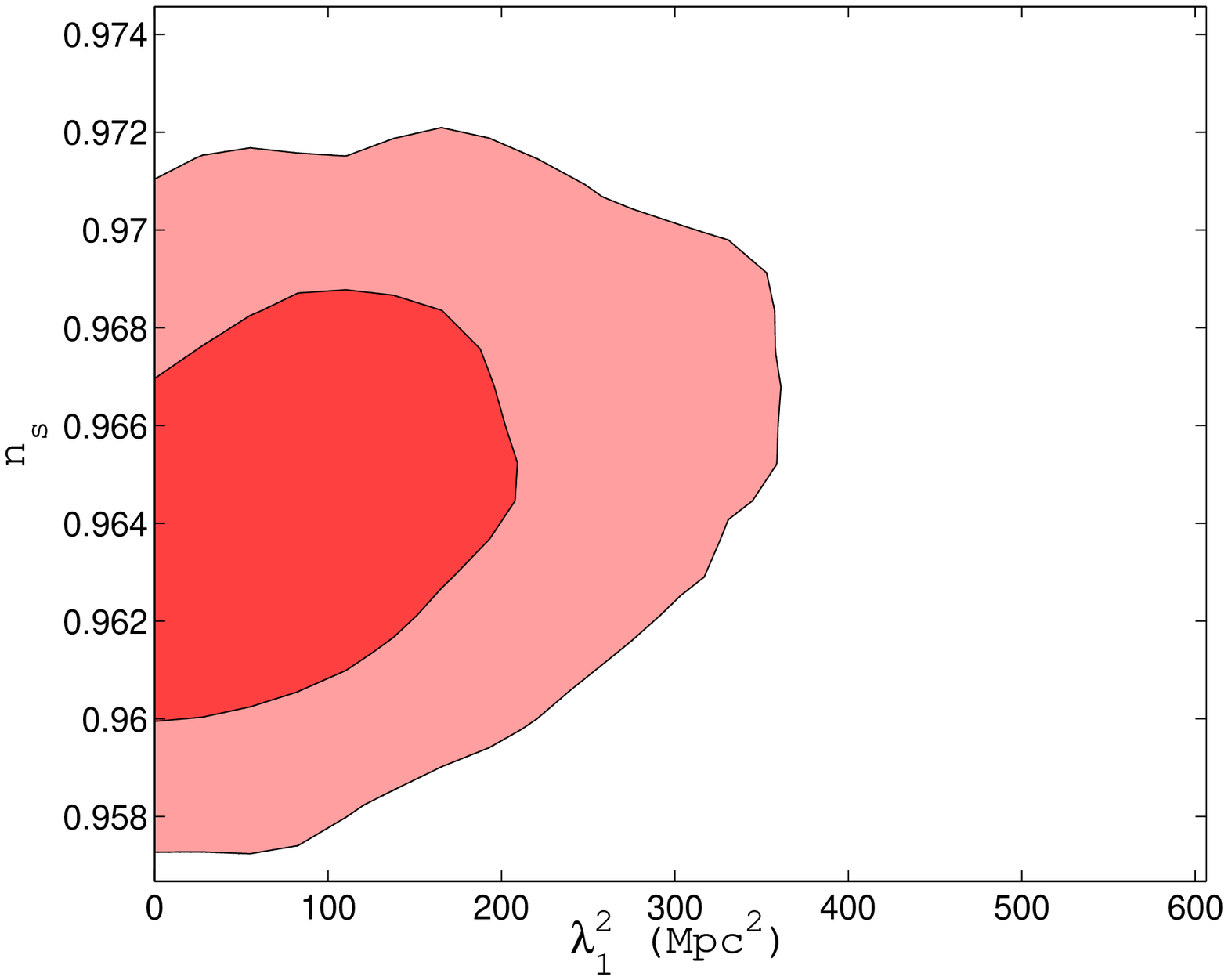,width=0.4\linewidth,clip=}\\
\end{tabular}
\caption{2-dimensional contour plots showing the degeneracies at $68
\%$ and $95 \%$ confidence levels for Planck on the left (blue
countours) and Planck+Euclid on the right (red
countours).Notice different scale for abscissae.}\label{countours}
\end{figure*}

In Table \ref{tab:results} we show the MCMC constraints at $68 \%$ c.l. for the \fr \ case
for Planck alone and Planck combined with Euclid. 
For this last case we also fit the data fixing $\lambda_1^2$ to $0$,
thus performing a standard analysis in a General Relativity framework, 
in order to show the importance of the degeneracies introduced by
$\lambda_1^2$ on the other cosmological parameters errors.
The parameters mostly correlated with modified
gravity are $H_0$ and $\Omega_ch^2$ (see also Figure
\ref{countours}) because these parameters strongly affect the
lensing convergence power spectrum as well as $\lambda_1^2$ through
$P(k,z)$. As expected in fact, when assuming general relativity we
find strong improvements on the errors on these parameters for the
combination Planck+Euclid in comparison to the varying $\lambda_1^2$ analysis. We note that the constraints
on the standard cosmological parameters are in good agreement with those showed in \cite{Refregier:2010ss}.

\begin{table}[!htb]
\begin{center}
\begin{tabular}{|l|c|c|c|cc|}
\hline

               & Planck& \multicolumn{2}{c|}{Planck+Euclid}  \\
\hline

Fiducial:                & $\lambda_1^2=0$ &   $\lambda_1^2=0$ &   $\lambda_1^2=0$ \\
Model:                   & varying $\lambda_1^2$ & varying  $\lambda_1^2$ & fixed $\lambda_1^2$\\
Parameter &&&\\
\hline
$\Delta{(\Omega_bh^2)}$ & $0.00013$             & $0.00011$        &  $0.00010$        \\
$\Delta{(\Omega_ch^2)}$ & $0.0010$              & $0.00073$        &  $0.00057$         \\
$\Delta{(\theta_s)}$    & $0.00027$             & $0.00025$         &  $0.00023$       \\
$\Delta{(\tau)}$        & $0.0041$              & $0.0030$         &  $0.0026$          \\
$\Delta{(n_s)}$         & $0.0031$              & $0.0029$           &  $0.0027$        \\
$\Delta{(\log[10^{10} A_s])}$ & $0.013$         & $0.0091$           &  $0.0091$        \\
$\Delta{(H_0)}$         & $0.50$                & $0.38$          &  $0.29$        \\
$\Delta{(\Omega_\Lambda)}$    & $0.0050$        & $0.0040$         &  $0.0031$         \\
$\lambda_1^2$(\rm Mpc$^2$) & $<2.42\times10^4$               & $<2.9\times10^2$ &   $-$ \\
\hline
\end{tabular}
\caption{$68 \%$ c.l. errors on cosmological parameters. Upper
limits on $\lambda_1^2$ are $95\%$ c.l. constraints.\\In the third
column we show constraints on the cosmological parameters when
fitting the data assuming general relativity, i.e. fixing
$\lambda_1^2=0$ \rm {Mpc}$^{2}$.} \label{tab:results}
\end{center}
\end{table}

In Figure \ref{countours} we show the $68\%$ and $95\%$ confidence
level 2-D likelihood contour plots in the $\Omega_m-\lambda_1^2$,
$H_0-\lambda_1^2$ and $n_s-\lambda_1^2$ planes, for Planck on the
left (blue) and Planck+Euclid on the right (red). As one can see the
inclusion of Euclid data can improve constraints on the standard
cosmological parameters from a $10\%$ to a $30\%$, with the most
important improvements on the dark matter physical density and the
Hubble parameter to which the weak lensing is of course very
sensitive as showed by Eq.~(\ref{convergencegalaxy}) and (\ref{weight}). 
Concerning modified gravity, Euclid data are
decisive to constrain $\lambda_1^2$, improving of two order of
magnitude the $95\%$ c.l. upper limit, thanks to the characteristic
effect of the modified gravity on the growth of
structures.\\

\begin{table}[!htb]\footnotesize
\begin{center}
\begin{tabular}{|l|c|c|c|}
\hline
& Planck+Euclid & Planck+Euclid & Fiducial values\\
\hline
Model: &$\lambda_1^2=0$& varying $\lambda_1^2$& \\
Parameter & & &\\
\hline
$\Omega_bh^2$       &  $0.022326\pm0.000096$ & $0.02259\pm0.00012$ & $0.02258$\\
$\Omega_ch^2$       &  $0.1126\pm0.00055$    & $0.11030\pm0.00083$ & $0.1109$\\
$\theta_s$          &  $1.0392\pm0.00023$    & $1.0395\pm0.00025$  & $1.0396$\\
$\tau$              &  $0.0775\pm0.0024$     & $0.08731\pm0.0029$   & $0.088$\\
$n_s$               &  $0.9592\pm0.0027$     & $0.9636\pm0.0029$   & $0.963$\\
$H_0$               &  $69.94\pm0.27$        & $71.20\pm0.42$      & $71.0$\\
$\Omega_\Lambda$    &  $0.724\pm0.003$       & $0.738\pm0.005$     & $0.735$\\
$\sigma_8$          &  $0.8034\pm0.0008$     & $0.8245\pm0.0039$   & $0.8239$\\
\hline
\end{tabular}
\caption{best fit value and $68 \%$ c.l. errors on cosmological
parameters for the case with a fiducial model $\lambda_1^2=300$
fitted with a $\Lambda$CDM model where $\lambda_1^2=0$ is assumed.}
\label{tab:shift}
\end{center}
\end{table}

\begin{figure*}[h!]
\centering
\begin{tabular}{cc}
\epsfig{file=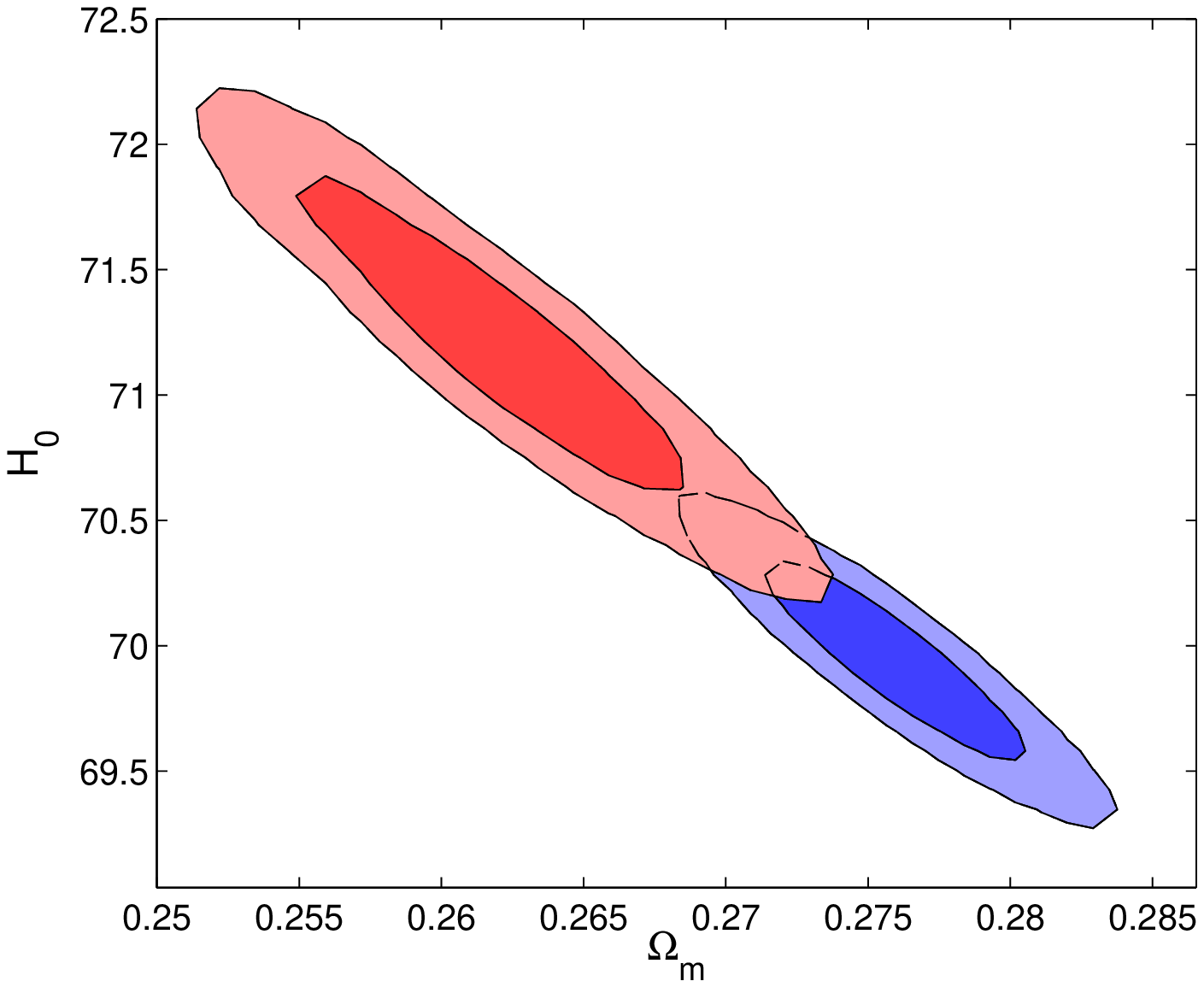,width=0.4\linewidth,clip=} &
\epsfig{file=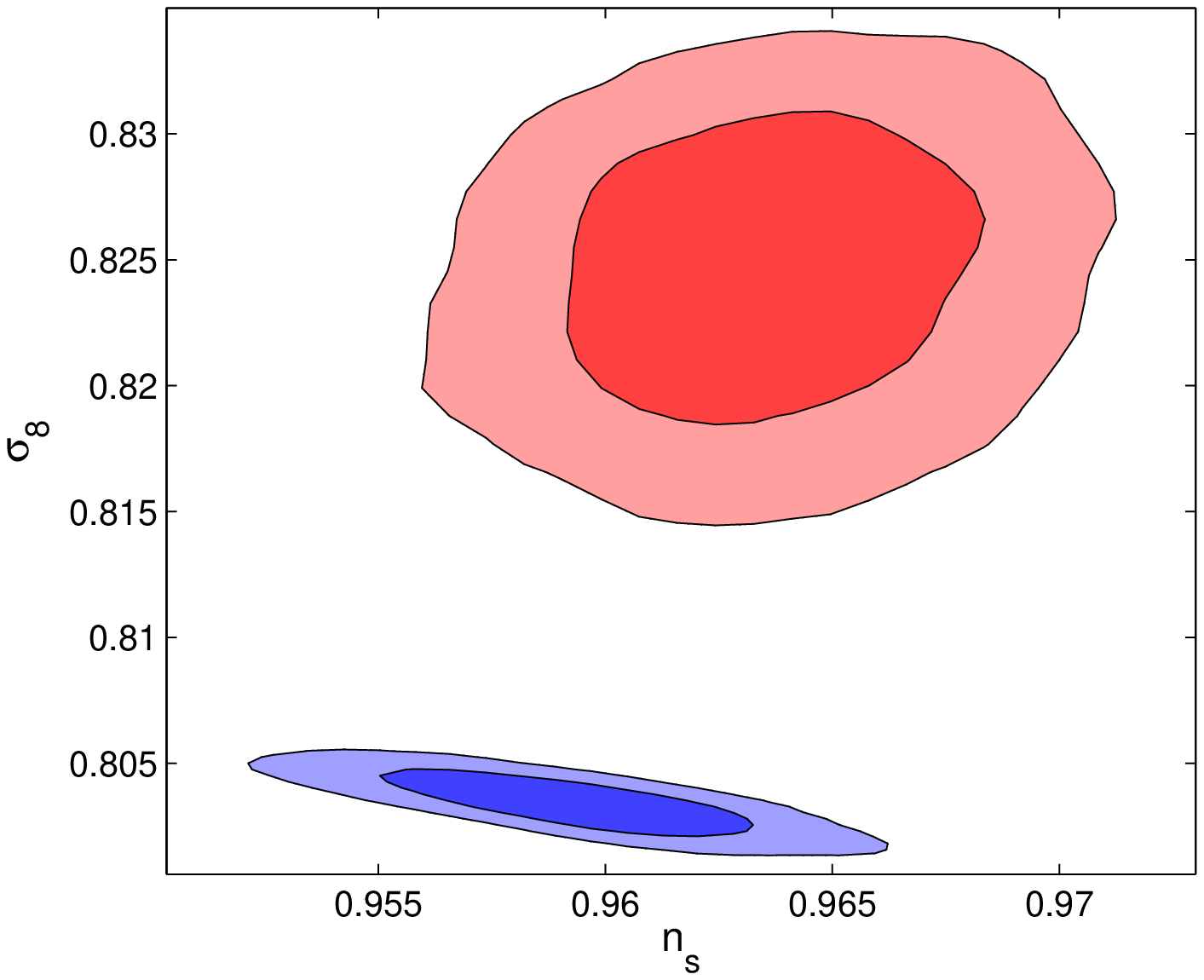,width=0.4\linewidth,clip=}\\
\end{tabular}
\caption{2-dimensional contour plots showing the degeneracies at $68
\%$ and $95 \%$ confidence levels for Planck+Euclid assuming a 	\fr \  
fiducial cosmology with $\lambda_1^2=300$\rm{Mpc}$^2$ considering an analysis 
with $\lambda_1^2$ fixed to $0$ (blue countours) or allowing it to vary (red
countours).}\label{shift}
\end{figure*}

Moreover, when analyzing the $f(R)$ mock datasets with
$\lambda_1^2=300$ \rm {Mpc}$^{2}$ as fiducial model, assuming 
$\lambda_1^2=0$ \rm {Mpc}$^{2}$ we
found a consistent bias in the recovered best fit value of the
cosmological parameters due to the degeneracies between
$\lambda_1^2$ and the other parameters. As it can be seen from the
comparison of Figures \ref{countours} and Figures \ref{shift} and from table
\ref{tab:shift} the shift in the best fit values is, as expected,
along the degeneracy direction of the parameters with $\lambda_1^2$,
for example for $n_s$, $H_0$ and $\Omega_m$. These results show that
for an even small modified gravity, the best fit values recovered by
wrongly assuming general relativity are more than $68 \%$ c.l. (for
some parameters at more than $95 \%$ c.l.) away from the correct fiducial
values, and may cause an underestimation of $n_s$ and $H_0$ and an
overestimation of $\sigma_8$ and $\Omega_m$. More generally, as
shown in table \ref{tab:shift}, all parameters are affected.\\ We
conclude, hence, that a future analysis of so high precision data
from Euclid and Planck will necessarily require to allow for
possible deviations from general relativity, in order to not bias
the best fit value of the cosmological parameters.\\

We also perform an analysis allowing $\beta_1$ and $s$ to vary;
in this way we can constrain not only \fr theories but also more general
scalar-tensor models, adding to the standard parameter set the time variation of 
the new gravitational interaction $s$ and the coupling with matter $\beta_1$.\\
We perform this analysis assuming as a fiducial model a  
\fr \ theory with $\lambda_1^2=3.0\times10^4$ \rm {Mpc}$^{2}$ and $\beta_1=4/3$.

\begin{table*}[!htb]
\begin{center}
\begin{tabular}{|l|c|c|c|}
\hline

               &  Planck & Planck+Euclid \\
Fiducial:                &   $\lambda_1^2=3.0\times10^4$ & $\lambda_1^2=3.0\times10^4$ \\
Parameter &\phantom{aaaaaaaaaa}&\\
\hline
$\Delta{(\Omega_bh^2)}$       & $0.00013$    & $0.00011$ \\
$\Delta{(\Omega_ch^2)}$       & $0.0011$     & $0.00082$ \\
$\Delta{(\theta_s)}$          & $0.00026$    & $0.00025$ \\
$\Delta{(\tau)}$              & $0.0043$     & $0.0040$  \\
$\Delta{(n_s)}$               & $0.0033$     & $0.0029$  \\
$\Delta{(\log[10^{10} A_s])}$ & $0.014$      & $0.011$   \\
$\Delta{(H_0)}$               & $0.54$       & $0.40$    \\
$\Delta{(\Omega_\Lambda)}$    & $0.0060$     & $0.0045$  \\
$\Delta{(\beta_1)}$           & $0.13$       & $0.038$   \\
$\lambda_1^2$                 & unconstrained & unconstrained \\
$s$                           & unconstrained & unconstrained \\
\hline
\end{tabular}
\caption{$68 \%$ c.l. errors on cosmological parameters and $\beta_1$.
We do not show limits on $\lambda_1^2$ and $s$ because this 
kind of analysis does not allow to constrain them (see text).} \label{tab:scalar}
\end{center}
\end{table*}

In Table \ref{tab:scalar} we report the $68 \%$ c.l. errors on the standard cosmological parameters, plus
the coupling parameter $\beta_1$. Performing a linear analysis, with a fiducial value of $\lambda_1^2=3\times10^4$, 
we obtain constraints on $\beta_1$ with  
$\Delta{(\beta_1)}=0.038$ at $68 \%$ c.l. and therefore potentially discriminating between modified gravity models and 
excluding the $\beta_1=1$ case (corresponding to the standard 
$\Lambda$CDM model) at more than $5-\sigma$ from a combination of Planck+Euclid data (only $2-\sigma$ for Planck alone).\\
The strong correlation present between $\beta_1$ and $\lambda_1^2$ (see eq. \ref{par_mu_old}) 
implies that, choosing a lower $\lambda_1^2$ fiducial value for a \fr \  model, the same variation of $\beta_1$ brings to 
smaller modifications of CMB power spectra and therefore we can expect weaker bounds on the coupling parameter.
In order to verify this behaviour we made three analysis fixing $s=4$ and choosing three different
fiducial values for $\lambda_1^2$: $3\times10^2$, $3\times10^3$ and $3\times10^4$ \rm {Mpc}$^{2}$.
The respectively obtained $\beta_1$ $68 \%$ c.l. errors are $0.11$, $0.052$ and $0.035$, confirming the decreasing expected accuracy
on $\beta_1$ for smaller fiducial values of $\lambda_1^2$.\\

The future constraints presented in this paper are obtained using a MCMC approach. Since most of the forecasts present in literature on \fr \ theories are obtained using a Fisher matrix analysis, it is useful to compare
our results with those predicted by a Fisher Matrix approach.
We therefore perform a Fisher Matrix analysis for Planck and Planck+Euclid 
(see \cite{DeBernardis:2008ys,Hannestad:2006as,Bond:1997wr}) assuming a \Lam fiducial model and we compare the results with those in Table \ref{tab:results}.\\
We find that for Planck alone the error on \lam is underestimated by a factor $\sim 3$
while the error is closer to the MCMC result for the Planck+Euclid case (underestimated by a factor $\sim 1.2$).

\section{Conclusions} \label{sec:vii}

In this paper we forecasted the ability of future weak lensing surveys as Euclid to
constrain modified gravity. We restricted our analysis to models that could mimic a cosmological
constant in the expansion of the Universe and can therefore be discriminated by only looking
at the growth of perturbations.
We have found that Euclid could improve the constraints on these models by nearly two order
of magnitudes respect to the constraints achievable by the Planck CMB satellite alone.
We have also discussed the degeneracies among the parameters and we found that neglecting the
possibility of modified gravity can strongly affect the constraints from Euclid on parameters as
the Hubble constant $H_0$, $\Omega_m$ and the amplitude of r.m.s. fluctuations $\sigma_8$. In this paper we found that, considering more general expansion histories, would further relax our constraints
and increase the degeneracies between the parameters. However other observables can be
considered as Baryonic Acoustic Oscillation and luminosity distances of high redshift supernovae
to further probe the value of $w$ and its redshift dependence.

\section{Acknowledgments}

It is a pleasure to thank Adam Amara and Luca Amendola 
for useful comments and suggestions. 
We also thank Gong-Bo Zhao for the latest version of the MGCAMB code.
Support was given by the Italian Space Agency through 
the ASI contracts "Euclid- IC" (I/031/10/0)


\begin{thebibliography}{srt}


\bibitem{Tsujikawa:2010sc}
  S.~Tsujikawa,
  arXiv:1004.1493 [astro-ph.CO].

\bibitem{Silvestri:2009hh}
  A.~Silvestri and M.~Trodden,
  Rept.\ Prog.\ Phys.\  {\bf 72} (2009) 096901
  [arXiv:0904.0024 [astro-ph.CO]].

\bibitem{Serra:2009yp}
  P.~Serra, A.~Cooray, D.~E.~Holz, A.~Melchiorri, S.~Pandolfi and D.~Sarkar,
  Phys.\ Rev.\  D {\bf 80} (2009) 121302
  [arXiv:0908.3186 [astro-ph.CO]].

\bibitem{Padmanabhan:2002ji}
  T.~Padmanabhan,
  Phys.\ Rept.\  {\bf 380} (2003) 235
  [arXiv:hep-th/0212290].

\bibitem{Peebles:2002gy}
  P.~J.~E.~Peebles and B.~Ratra,
  Rev.\ Mod.\ Phys.\  {\bf 75} (2003) 559
  [arXiv:astro-ph/0207347].

\bibitem{Laszlo:2007td}
  I.~Laszlo and R.~Bean,
  Phys.\ Rev.\  D {\bf 77} (2008) 024048
  [arXiv:0709.0307 [astro-ph]].

\bibitem{Amendola:2007rr}
  L.~Amendola, M.~Kunz and D.~Sapone,
  JCAP {\bf 0804} (2008) 013
  [arXiv:0704.2421 [astro-ph]].

\bibitem{Linder:2007hg}
  E.~V.~Linder and R.~N.~Cahn,
  Astropart.\ Phys.\  {\bf 28} (2007) 481
  [arXiv:astro-ph/0701317].

\bibitem{Bartelmann:1999yn}
  M.~Bartelmann and P.~Schneider,
  Phys.\ Rept.\  {\bf 340} (2001) 291
  [arXiv:astro-ph/9912508].

\bibitem{Refregier:2003ct}
  A.~Refregier,
  Ann.\ Rev.\ Astron.\ Astrophys.\  {\bf 41} (2003) 645
  [arXiv:astro-ph/0307212].

\bibitem{VanWaerbeke:2003uq}
  L.~Van Waerbeke and Y.~Mellier,
  arXiv:astro-ph/0305089.

\bibitem{Daniel:2010ky}
  S.~F.~Daniel, E.~V.~Linder, T.~L.~Smith, R.~R.~Caldwell, A.~Cooray, A.~Leauthaud and L.~Lombriser,
  Phys.\ Rev.\  D {\bf 81} (2010) 123508
  [arXiv:1002.1962 [astro-ph.CO]].

\bibitem{Lombriser:2010mp}
  L.~Lombriser, A.~Slosar, U.~Seljak and W.~Hu,
  arXiv:1003.3009 [astro-ph.CO].

\bibitem{Bean:2010zq}
  R.~Bean and M.~Tangmatitham,
  Phys.\ Rev.\  D {\bf 81} (2010) 083534
  [arXiv:1002.4197 [astro-ph.CO]].

\bibitem{Zhao:2010dz}
  G.~B.~Zhao {\it et al.},
  Phys.\ Rev.\  D {\bf 81} (2010) 103510
  [arXiv:1003.0001 [astro-ph.CO]].

\bibitem{Daniel:2008et}
  S.~F.~Daniel, R.~R.~Caldwell, A.~Cooray and A.~Melchiorri,
  Phys.\ Rev.\  D {\bf 77} (2008) 103513
  [arXiv:0802.1068 [astro-ph]].

\bibitem{Daniel:2009kr}
  S.~F.~Daniel {\it et al.},
  Phys.\ Rev.\  D {\bf 80} (2009) 023532
  [arXiv:0901.0919 [astro-ph.CO]].

\bibitem{Refregier:2010ss}
  A.~Refregier, A.~Amara, T.~D.~Kitching, A.~Rassat, R.~Scaramella, J.~Weller and f.~t.~E.~Consortium,
  arXiv:1001.0061 [astro-ph.IM].

\bibitem{Refregier:2006vt}
  A.~Refregier {\it et al.},
  arXiv:astro-ph/0610062.

\bibitem{Gehrels:2010fn}
  N.~Gehrels,
  arXiv:1008.4936 [astro-ph.CO].

\bibitem{Thomas:2008tp}
  S.~A.~Thomas, F.~B.~Abdalla and J.~Weller,
  Mon.\ Not.\ Roy.\ Astron.\ Soc.\  {\bf 395} (2009) 197
  [arXiv:0810.4863 [astro-ph]].

\bibitem{Xia:2009gb}
  J.~Q.~Xia,
  Phys.\ Rev.\  D {\bf 79} (2009) 103527
  [arXiv:0907.4860 [astro-ph.CO]].
\bibitem{DeBernardis:2008ys}
  F.~De Bernardis, L.~Pagano, P.~Serra, A.~Melchiorri and A.~Cooray,
  JCAP {\bf 0806} (2008) 013
  [arXiv:0804.1925 [astro-ph]].
\bibitem{Hannestad:2006as}
  S.~Hannestad, H.~Tu and Y.~Y.~Y.~Wong,
  JCAP {\bf 0606} (2006) 025
  [arXiv:astro-ph/0603019].
\bibitem{Bond:1997wr}
  J.~R.~Bond, G.~Efstathiou and M.~Tegmark,
  Mon.\ Not.\ Roy.\ Astron.\ Soc.\  {\bf 291} (1997) L33
  [arXiv:astro-ph/9702100].
\bibitem{Zhao:2008bn}
  G.~B.~Zhao, L.~Pogosian, A.~Silvestri and J.~Zylberberg,
  Phys.\ Rev.\  D {\bf 79} (2009) 083513
  [arXiv:0809.3791 [astro-ph]].

\bibitem{Heavens:2007ka}
  A.~F.~Heavens, T.~D.~Kitching and L.~Verde,
  Mon.\ Not.\ Roy.\ Astron.\ Soc.\  {\bf 380} (2007) 1029
  [arXiv:astro-ph/0703191].

\bibitem{Calabrese:2009tt}
  E.~Calabrese, A.~Cooray, M.~Martinelli, A.~Melchiorri, L.~Pagano, A.~Slosar and G.~F.~Smoot,
  Phys.\ Rev.\  D {\bf 80} (2009) 103516
  [arXiv:0908.1585 [astro-ph.CO]].

\bibitem{Serra:2009kp}
  P.~Serra, A.~Cooray, S.~F.~Daniel, R.~Caldwell and A.~Melchiorri,
  Phys.\ Rev.\  D {\bf 79} (2009) 101301
  [arXiv:0901.0917 [astro-ph.CO]].

\bibitem{Hu:2007nk}
  W.~Hu and I.~Sawicki,
  Phys.\ Rev.\  D {\bf 76} (2007) 064004
  [arXiv:0705.1158 [astro-ph]].

\bibitem{Bertschinger:2008zb}
  E.~Bertschinger and P.~Zukin,
  Phys.\ Rev.\  D {\bf 78} (2008) 024015
  [arXiv:0801.2431 [astro-ph]].

\bibitem{Khoury:2003aq}
  J.~Khoury and A.~Weltman,
  Phys.\ Rev.\ Lett.\  {\bf 93} (2004) 171104
  [arXiv:astro-ph/0309300].

\bibitem{Magnano:1993bd}
  G.~Magnano and L.~M.~Sokolowski,
  Phys.\ Rev.\  D {\bf 50} (1994) 5039
  [arXiv:gr-qc/9312008].

\bibitem{Giannantonio:2009gi}
  T.~Giannantonio, M.~Martinelli, A.~Silvestri and A.~Melchiorri,
  JCAP {\bf 1004} (2010) 030
  [arXiv:0909.2045 [astro-ph.CO]].

\bibitem{Munshi:2006fn}
  D.~Munshi, P.~Valageas, L.~Van Waerbeke and A.~Heavens,
  Phys.\ Rept.\  {\bf 462} (2008) 67
  [arXiv:astro-ph/0612667].

\bibitem{kai:1992}
N. Kaiser, Astrophys. J. 388 (1992) 272.

\bibitem{Kaiser:1996tp}
  N.~Kaiser,
  Astrophys.\ J.\  {\bf 498} (1998) 26
  [arXiv:astro-ph/9610120].


\bibitem{Perotto:2006rj}
  L.~Perotto, J.~Lesgourgues, S.~Hannestad, H.~Tu and Y.~Y.~Y.~Wong,
  JCAP {\bf 0610} (2006) 013
  [arXiv:astro-ph/0606227].

\bibitem{hirata:2003}
  C.~M.~Hirata and U.~Seljak,
  Phys.\ Rev.\  D {\bf 68} (2003) 083002
  [arXiv:astro-ph/0306354].


\bibitem{lensextr}
  T.~Okamoto and W.~Hu,
  Phys.\ Rev.\  D {\bf 67} (2003) 083002
  [arXiv:astro-ph/0301031].


\bibitem{Heavens:2003}
A. F. Heavens,
2003, MNRAS, {\bf 323}, 1327

\bibitem{Castro:2005}
P. G. Castro, A. F. Heavens, T. D. Kitching, 2005, Phys. Rev. D,
{\bf 72}, 3516

\bibitem{Heavens:2006}
A. F. Heavens,  T. D. Kitching, A. N. Taylor, 2006, MNRAS, {\bf
373}, 105

\bibitem{Kitching:2007}
T. D. Kitching, A. F. Heavens, A. N. Taylor, M. L. Brown, K.
Meisenheimer, C. Wolf, M. E. Gray, D. J. Bacon, 2007, MNRAS, {\bf
376}, 771


\bibitem{Hannestad:2006as}
  S.~Hannestad, H.~Tu and Y.~Y.~Y.~Wong,
  JCAP {\bf 0606} (2006) 025
  [arXiv:astro-ph/0603019].


\bibitem{art:Bacon} Bacon, D.; et al.; 2003; MNRAS, 363, 723-733

\bibitem{art:Massey} Massey R.; et al.; 2007, ApJS, 172, 239


\bibitem{art:Taylor} Taylor, A. N.; et al.; 2004, MNRAS, 353, 1176




\bibitem{Cooray:1999rv}
  A.~R.~Cooray,
  Astron.\ Astrophys.\  {\bf 348} (1999) 31
  [arXiv:astro-ph/9904246].

\bibitem{:2006uk}
    [Planck Collaboration],
  arXiv:astro-ph/0604069.

\bibitem{wmap7}
  E.~Komatsu {\it et al.},
  arXiv:1001.4538 [astro-ph.CO].




\bibitem{Smith:2002dz}
  R.~E.~Smith {\it et al.}  [The Virgo Consortium Collaboration],
  Mon.\ Not.\ Roy.\ Astron.\ Soc.\  {\bf 341} (2003) 1311
  [arXiv:astro-ph/0207664].




\bibitem{Lewis:2002ah}
A. Lewis and S. Bridle,
Phys.\ Rev.\ D {\bf 66}, 103511 (2002) (Available from
\texttt{http://cosmologist.info}.)


\bibitem{Refregier:2010ss}
  A.~Refregier, A.~Amara, T.~D.~Kitching, A.~Rassat, R.~Scaramella, J.~Weller and f.~t.~E.~Consortium,
  arXiv:1001.0061 [astro-ph.IM].


\bibitem{DeBernardis:2008ys}
  F.~De Bernardis, L.~Pagano, P.~Serra, A.~Melchiorri and A.~Cooray,
  JCAP {\bf 0806} (2008) 013
  [arXiv:0804.1925 [astro-ph]].
\bibitem{Hannestad:2006as}
  S.~Hannestad, H.~Tu and Y.~Y.~Y.~Wong,
  JCAP {\bf 0606} (2006) 025
  [arXiv:astro-ph/0603019].
\bibitem{Bond:1997wr}
  J.~R.~Bond, G.~Efstathiou and M.~Tegmark,
  Mon.\ Not.\ Roy.\ Astron.\ Soc.\  {\bf 291} (1997) L33
  [arXiv:astro-ph/9702100].


\end{thebibliography}
\end{document}